# Polarization-engineered GaN/InGaN/GaN tunnel diodes

Sriram Krishnamoorthy, [1,a)] Digbijoy N. Nath, [1] Fatih Akyol, [1] Pil Sung Park, [1] Michele Esposto, [1] Siddharth Rajan[1]

[1]Department of Electrical & Computer Engineering, The Ohio State University, Columbus, OH 43210

**Abstract**: We report on the design and demonstration of polarization-engineered GaN/InGaN/GaN tunnel junction diodes with high current density and low tunneling turn-on voltage. Wentzel-Kramers-Brillouin (WKB) calculations were used to model and design tunnel junctions with narrow bandgap InGaN-based barrier layers. N-polar p-GaN/In$_{0.33}$Ga$_{0.67}$N/n-GaN heterostructure tunnel diodes were grown using molecular beam epitaxy. Efficient zero bias tunneling turn-on with a high current density of 118 A/cm$^2$ at a reverse bias of 1V, reaching a maximum current density up to 9.2 kA/cm$^2$ were obtained. These results represent the highest current density reported in III-nitride tunnel junctions, and demonstrate the potential of III-nitride tunnel devices for a broad range of optoelectronic and electronic applications.

---

[a)] Author to whom correspondence should be addressed. Electronic mail: krishnas@ece.osu.edu
Tel: +1-614-688-8458

The phenomenon of inter-band tunneling across degenerately doped p-n junction, first reported by Esaki[1], has been exploited in a variety of device applications such as light emitting diodes (LEDs), laser diodes, multi junction solar cells[2,3], and more recently tunnel field effect transistors (TFETs)[4]. Tunnel junctions enable multi-junction solar cells due to their low resistance in reverse bias regime. In wide band gap devices such as III-nitride lasers and LEDs[5, 6], efficient tunnel junctions (TJ) may mitigate losses due to high resistance p-type layers and p-contacts. Tunneling based devices such as the TFETs are also promising for high performance and low-power computation as they offer a way to achieve sub threshold swing below the 60 mV/decade limit.

Extensive research has led to efficient inter-band tunneling based devices in material systems such as the III-As and SiGe[7]. While degenerately doped p-n tunnel junctions have been used to replace resistive p-GaN with n-GaN as the top contact layer for LED structures [8, 9], the III-nitride material system still lacks an efficient tunnel junction due to the large band gap and dopant solubility limits. For example, even with a very high doping of $10^{20}$ cm$^{-3}$, the depletion width of a standard p+/n+ GaN junction is 6 nm, but the large potential barrier due to higher band gap leads to high tunneling resistance and low current density.

The high spontaneous and piezoelectric polarization charge[10] along the c-axis of III-nitrides and other highly polar semiconductors provides a new design approach for tunneling structures. In a conventional tunnel junction, the degenerate doping reduces the space charge region thickness and creates band bending at the junction to align the valence and conduction bands on either side of the space charge region. The space charge dipole created by ionized donors and acceptors can be substituted with the fixed positive and negative polarization sheet charges at polar hetero-interfaces. In heterostructures of highly polar materials such as nitrides, polarization induced dipole can create significantly high electric fields resulting in large band bending over a small distance, thereby increasing the tunneling probability. This principle of polarization induced tunnel junction[11, 12] was demonstrated using AlN as the barrier material for inter-band tunneling in GaN. However, the large band gap of AlN reduces tunneling probability and the GaN/AlN/GaN devices have low current density and high tunneling resistance.

Tunneling probability across a potential barrier is determined by the barrier height and the thickness of the barrier material. Hence it is expected that for the same tunneling width, a lower band gap material will be a more efficient tunnel barrier between wide band gap materials. It may therefore be expected that $In_xGa_{1-x}N$ with its low band gap and a large piezoelectrically induced polarization when coherently strained to GaN, is an ideal candidate for efficient inter-band tunneling. In this letter, we investigate the potential of InGaN as a barrier material for efficient inter-band tunneling in GaN through calculations based on polarization and Wentzel-Kramers-Brillouin (WKB) approximation. We also demonstrate the first InGaN based tunnel junction grown by plasma assisted molecular beam epitaxy (PAMBE) with very high current density surpassing previous doping and GaN/AlN/GaN based tunnel diodes.

The equilibrium band diagram of the proposed GaN/InGaN/GaN tunnel junction is shown in Fig. 1. For zero bias interband tunneling to occur, the InGaN composition and thickness are chosen such that the polarization induced band bending aligns the conduction and valence band on either side of the sandwiched InGaN layer. GaN is degenerately doped to reduce the depletion region thickness at the GaN/InGaN interface, and hence the major barrier to tunneling is the InGaN layer. Under reverse bias, the electrons in the valence band of p-GaN tunnel across the p-depletion region (intra valence band), InGaN (inter-band), and n-depletion region (intra conduction band), entering the conduction band of n-GaN as shown in the inset of Fig.1. The equilibrium band bending $\Phi$ due to $In_xGa_{1-x}N$ layer of thickness 't' is $\Phi(x) = q\,\sigma(x)\,t\,/\,\varepsilon(x)$, where $\varepsilon(x)$ is the permittivity of $In_xGa_{1-x}N$ and $\sigma(x)$ is the fixed polarization induced charge density at the $GaN/In_xGa_{1-x}N$ interface[13]. For a critical InGaN layer thickness, $t_{cr}$, the equilibrium potential drop in the InGaN layer equals the band gap of the InGaN layer, $E_{g,InGaN}$. If $t < t_{cr}$, tunneling cannot occur at zero bias since conduction and valence band extrema on either side are not aligned. With $t > t_{cr}$, tunneling probability reduces due to the increased thickness of the barrier. Fig.2 shows the variation of the critical thickness $t_{cr}$ with the InN mole fraction in InGaN. The critical thickness, $t_{cr}$, decreases with increasing In composition due to the combined effects of increasing polarization charge and decreasing band gap. For the critical thickness $t_{cr}$ the potential barrier V (t) seen by an electron is a triangular barrier of width $t_{cr}$ and maximum height of $E_{g,InGaN}/q$ as illustrated in the inset of Fig.2. The tunneling probability p across the barrier, for a critical thickness $t_{cr}$ is evaluated by using Wentzel-Kramers-Brillouin (WKB) approximation[14] given by

$$p(t_{cr}) = \exp\left(-\int_0^{t_{cr}} \sqrt{\frac{2m^* E_{g,InGaN} t}{\hbar^2 q t_{cr}}} dt\right) \quad (1)$$

where $m^*$ is the effective mass of InGaN.

The tunneling probability 'p' calculated using Eq. (1), is found to increase as the In composition in the barrier material is increased (Fig. 2). Also, the tunneling probability for a GaN/ AlN(2.6nm)/ GaN structure, calculated to be ~ $10^{-11}$, is shown for comparison as a dotted line in Fig.2. This is the optimal thickness and composition for an AlGaN-based tunnel junction. It can be observed that for higher In composition, the tunneling probability of the GaN/InGaN/GaN TJ is several orders of magnitude higher than that for GaN/AlN/GaN TJ. It should be noted that only the tunneling across the InGaN barrier is considered in calculating the probability, and this component increases as the In composition in increased. However, for very high In compositions at which the band offsets are of the order of bandgap of InGaN, tunneling barrier due to the depletion regions at the GaN/InGaN heterojunctions can no longer be neglected and may play a significant role in reducing the tunneling probability.

To demonstrate the potential advantage in using a GaN/InGaN/GaN structure, an InGaN TJ sample was grown on a Lumilog[15] N-polar free standing LED quality GaN template (dislocation density ~ $10^8$ cm$^{-2}$) by PAMBE in a Veeco Gen 930 system. The epitaxial structure of the device is shown in the inset of Fig.3. The N-polar orientation is used in this work, as higher In incorporation can be achieved at a given growth temperature compared to the conventional Ga-polar orientation.[16,17] N-polar InGaN was grown on 100nm thick Si doped GaN ($N_D$ ~ 5 X $10^{18}$ cm$^{-3}$) using the conditions and growth model developed earlier[17]. The growth model enables us to grow a specified thickness and composition of InGaN while taking into account the composition and temperature dependent decomposition rate. InGaN was grown at a high substrate temperature of $600^0$C for superior quality. The InGaN layer was capped with 100nm of p GaN ($N_A \sim 1 \times 10^{19}$ cm$^{-3}$). The thickness and composition of the InGaN layer were assessed to be 6.4 nm and 33.5 % respectively from ω-2θ triple-axis scans (not shown here) using a BEDE high resolution X-ray diffractometer, indicating good agreement with the growth model[18]. A reference p-n junction with identical doping profiles and growth conditions but without an InGaN layer was grown as a control sample. Both the samples exhibited step flow growth morphology as seen using atomic force microscopy (not shown here). Ni/Au (20/150 nm) and Ti/Au (20/100 nm) stacks were evaporated using standard optical lithography on the p-GaN and n-GaN layers respectively.

Fig. 4 shows the typical I-V characteristic of the InGaN TJ device (30 x 30 μm$^2$) and a reference pn junction. In the forward bias condition, the TJ and pn junction sample had similar current density. The turn-on voltage of the TJ was lower, which we attribute to the presence of a lower bandgap InGaN quantum well. In the reverse direction, the pn junction shows reverse leakage current which is lower than the forward current. The TJ shows much higher reverse current than in the forward direction, which is a typical behavior of a backward diode. In addition, the reverse tunneling current in the TJ was several orders of magnitude higher than in the pn junction.

Two regimes were observed in the reverse I-V characteristics for the GaN/InGaN/GaN TJ. In the low reverse bias regime, a sharp increase in reverse current was observed (~ 70-130 mV/decade) from zero bias (inset to Fig. 3) indicating the onset of tunneling in the device. Such a low turn on voltage would be an ideal candidate to connect devices in series, especially in the case of multi-junction solar cells. At a reverse voltage of 1 V, a current density of 118 A/cm$^2$ was obtained. In comparison, the reference pn junction sample had very low reverse current (5 orders of magnitude lower at -1 V) as expected due to the thicker barrier in the absence of polarization induced field. At higher current density levels (> 100 A/cm$^2$), the differential resistance increased significantly. Further analysis is needed to understand the origin of this behavior, but it may be attributed partly to increased series and contact resistances and self-heating. The maximum observed current density in reverse bias, 9.2 kA/ cm$^2$, is the highest current density reported for III-nitride tunnel diodes.

While the results described here on GaN/InGaN/GaN TJs demonstrate clearly the promise of this technology, further improvements in the device characteristics can be achieved by using higher composition InGaN and optimizing barrier thickness. Further sophistication in design, such as graded GaN/InGaN interfaces to eliminate abrupt depletion barriers, asymmetric junctions, and the use of quarternary alloys may enable lower resistance and higher tunneling current. A complete theoretical understanding and modeling of the intra-band and inter-band tunneling transport in this structure will enable better design of these devices. In addition, band structure effects are expected to play an important role at high current density, and a more detailed calculation incorporating these may provide insight[19]. The concept of enhancing tunneling probabilities by using a narrow gap material as a tunnel barrier demonstrated here may be used as the basis for other III-nitride alloys such as AlGaN and AlN, as well as other wide band gap material systems that have significant polarization charge (such as ZnO).

In conclusion, we have demonstrated the promise of polarization engineering using narrow bandgap layers as a barrier material for interband tunneling in III-nitride devices. WKB calculations were used to model and design GaN/InGaN/GaN tunnel junctions. Extremely low tunneling resistance, zero bias turn-on, and and high reverse current density (100 A/ cm$^2$ at -1 V, $J_{max}$ = 9.2 kA/cm$^2$) were achieved with a GaN/In$_{0.33}$Ga$_{0.67}$N/GaN tunnel junction. The tunnel junction designs demonstrated here will enable the incorporation of tunnel junctions in several technologically relevant III-nitride devices such as LEDs, lasers, and solar cells, and provide a pathway to novel device structures such as tunnel FETs.

We would like to acknowledge funding from ONR (Program manager: Paul Maki) and OSU Institute for Materials Research (IMR).

[19] M. F. Shubert, Phys. Rev. B **81**, 035303 (2010).

**Figure captions**:

**Figure 1**: Equilibrium energy band diagram of GaN/In$_{0.33}$Ga$_{0.67}$N/GaN zero bias inter-band tunnel junction. **Inset:** Band diagram at reverse bias showing inter-band tunneling.

**Figure 2**: (Color Online) Critical tunnel junction thickness t$_{cr}$ in order to achieve zero bias inter-band tunneling is shown as a function of InN mole fraction in InGaN. Tunneling probability computed as a function of InN mole fraction in InGaN for a barrier thickness of t$_{cr}$ is shown. Dotted line indicates a tunneling probability of $10^{-11}$ for a GaN/ AlN (2.6nm)/ GaN structure. **Inset:** Potential profile of the barrier for electron tunneling used for calculation of p

**Figure 3**: (Color Online) Log J-V characteristics of the GaN/In$_{0.33}$Ga$_{0.67}$N/GaN TJ (solid line) and standard p+/n+ junction (dotted line) and. **Inset (top right):** Epitaxial structure of the InGaN TJ sample. **Inset (bottom left):** Linear J-V characteristics of the GaN/In$_{0.33}$Ga$_{0.67}$N/GaN TJ (solid line) and reference p+/n+ junction (dotted line).

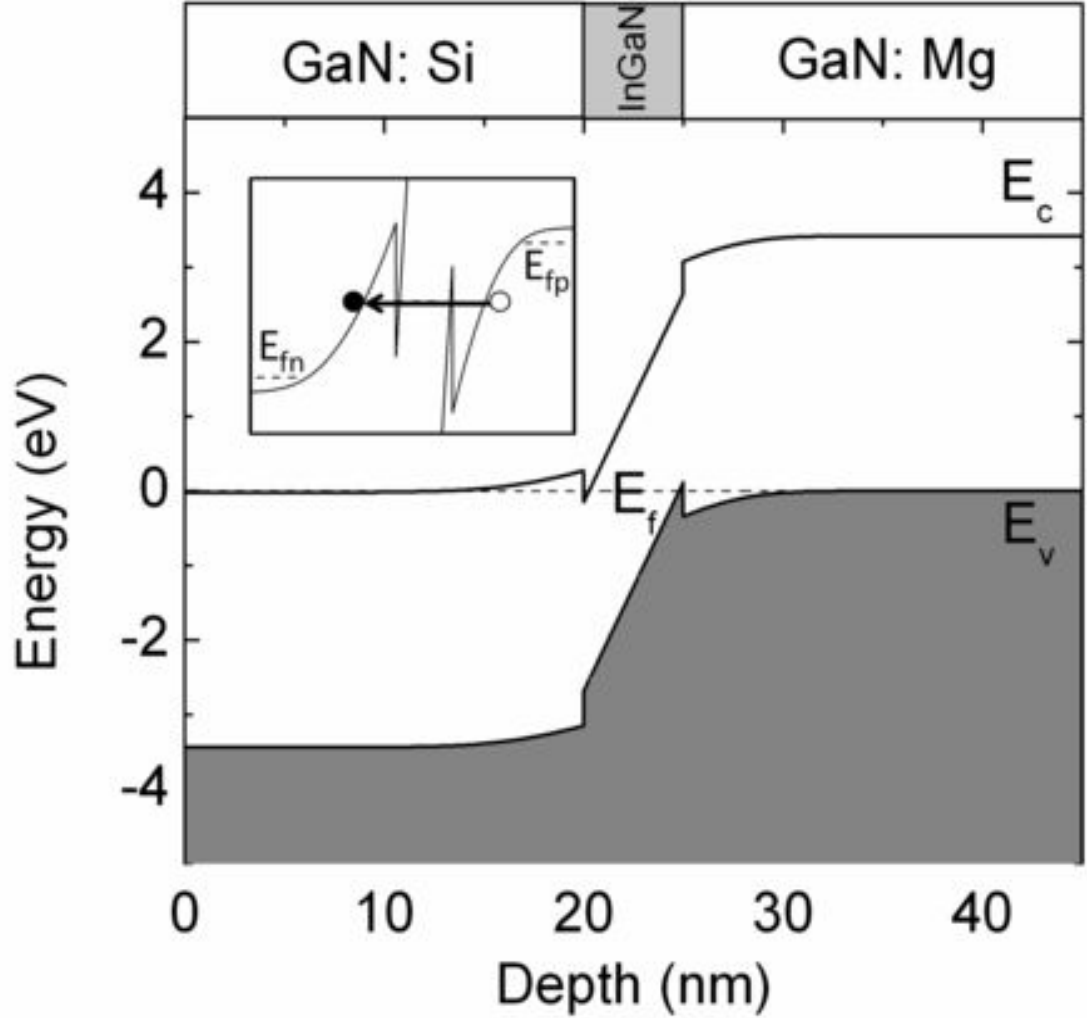

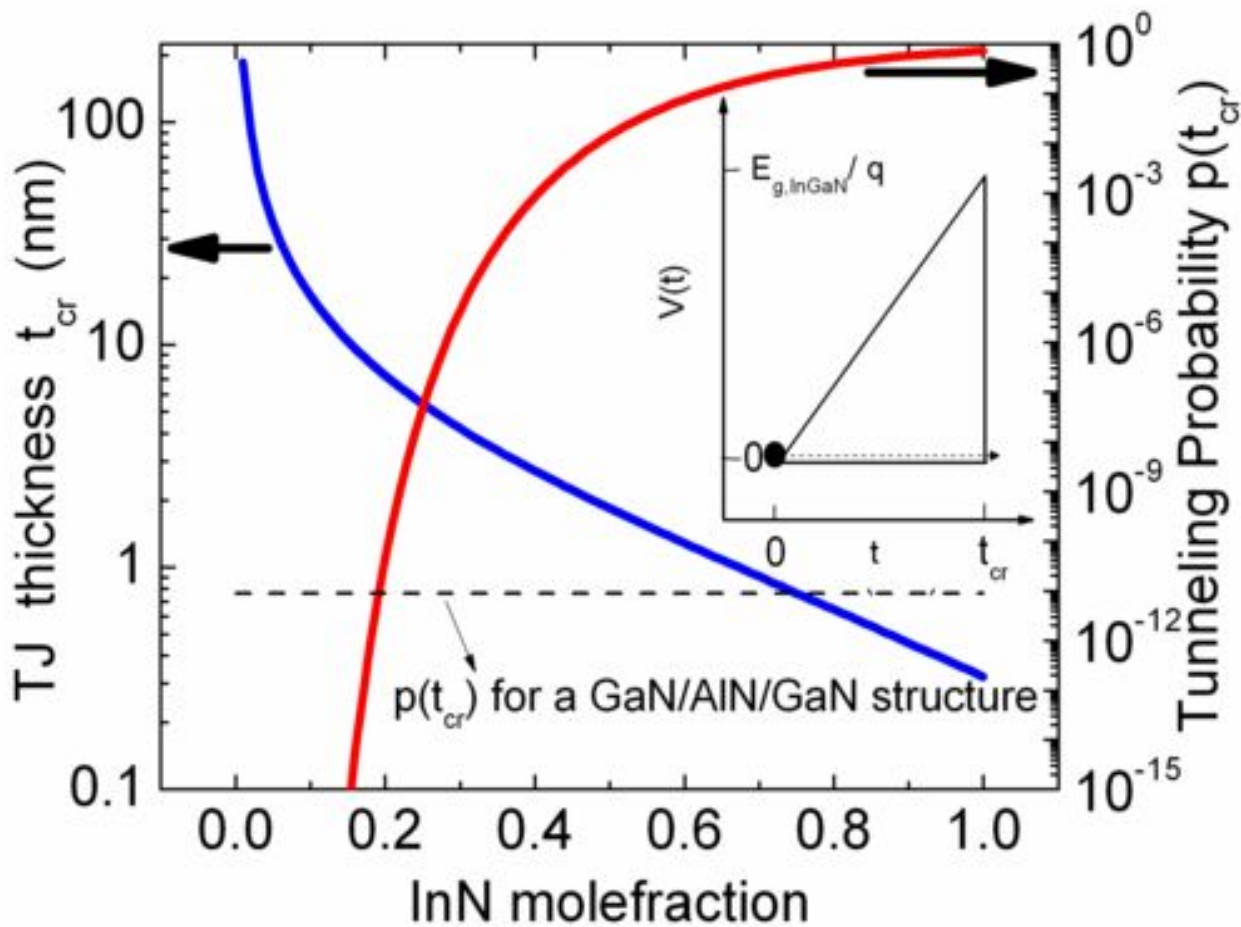

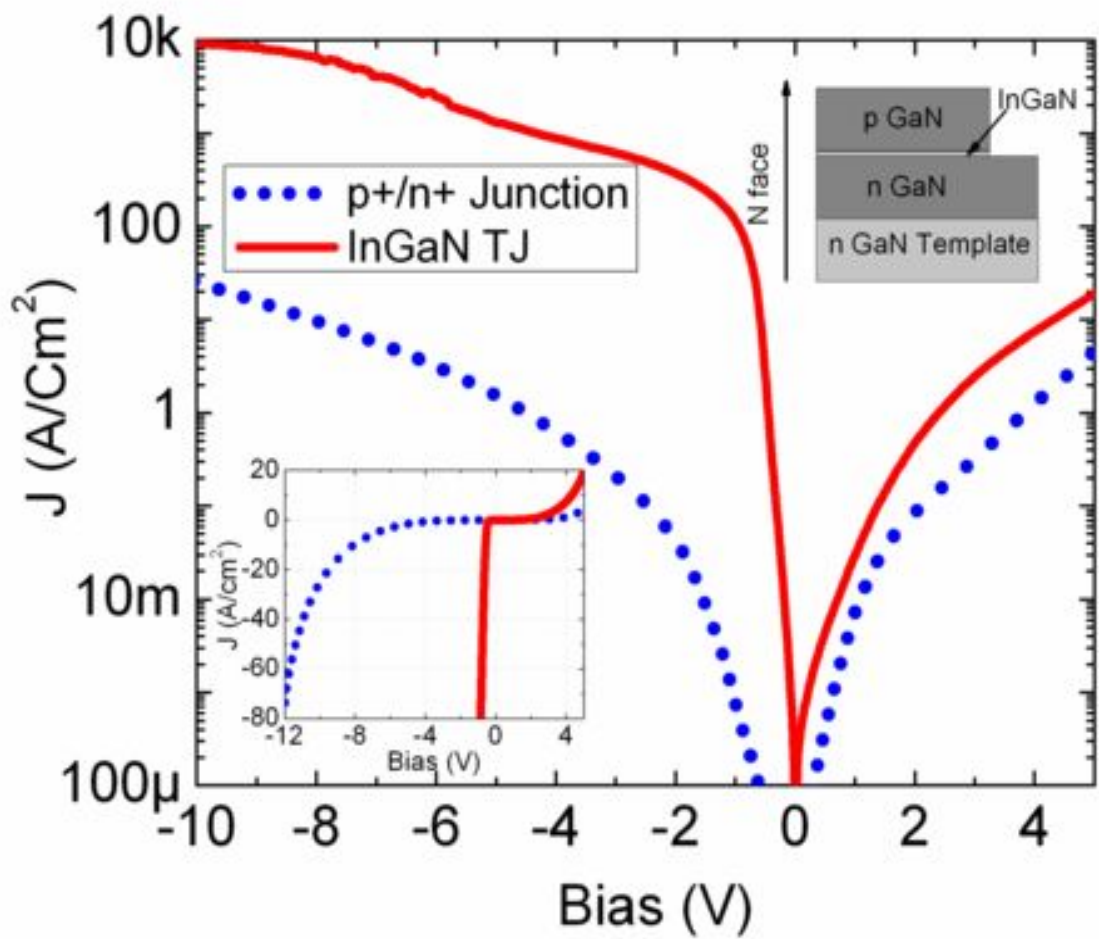